# Data Transmission and Communication via Electrolytic Flow Channel

Khoi Ly*, Chong-Chan Kim*, Robert Shepherd

*The authors contribute equally to this archive document

Khoi Ly, Ph.D. kdl54@cornell.edu, Sibley School of Mechanical and Aerospace Engineering, Cornell University

Chong-Chan Kim, Ph.D ck775@cornell.edu, Sibley School of Mechanical and Aerospace Engineering, Cornell University

Robert Shepherd, Ph.D rfs247@cornell.edu, Sibley School of Mechanical and Aerospace Engineering, Cornell University

**Abstract:** As an alternative approach to ionic data transmission with hydrogel as substrate, this work explores the possible applications of liquid electrolyte filling cavity of a stretchable, flexible elastomeric tubing, which is the primary ingredient used in redox flow battery systems. While hydrogel-based ionic impedance characterization and its data communication capability have been well studied, the multifunctional use of redox flow battery's electrolyte for data communication, in addition to powering, is novel, especially in the context of soft robotics. This work also describes a simple signal conditioning technique that addresses the signal decaying problem due to long-distance ionic data transmission. Finally, we describe the prototypical design of a decentralized data communication between two, or possibly more, systems of power and control. This work presents the concept of a decentralized control of a robot's hydraulic actuation system, that will allow for the system's functional robustness in the event of one of the two, or possibly more, modules lose power or become severed from the robot's body.

## I. Introduction

One of the largest fundamental challenges in soft robots that mimic animals' functions is the design of stretchable, compliant nervous systems, particularly for sensorimotor feedback, perception, and control. Traditionally, robots use rigid, hard metallic wires for sensing and controlling various components, whereas animals use axons for sending and receiving sensorimotor signals. The key difference between the metallic wires and axons is how the data is transmitted. Metallic wires contain free electrons that can easily move between atoms, whereas the axons transmit data based on action potentials caused by ionic concentration gradients. From literature, engineered stretchable ionic conductors integrate the two functions – stretchability and conductivity – at the molecular scale. A gel consists of a polymer network and a solvent (e.g., water or ionic liquids). The polymer network makes the gel a soft elastic solid, and the solvent makes the gel a fast ionic conductor. Thanks to high compliance while retaining excellent electrical conductivity, hydrogel based ionic conductors facilitate many applications in prosthesis (Sarwar 2017), electroceuticals (Svirskis 2010), human-machine interfaces (Sheng 2019), and soft robotics (Rumley 2023).

As an alternative approach to ionic data transmission with hydrogel as substrate, this work explores the possible applications of liquid electrolyte filling cavity of a stretchable, flexible elastomeric tubing, which is the primary ingredient used in redox flow battery systems. While hydrogel-based ionic impedance characterization and its data communication capability have been well studied (Yang 2015; Shi 2020; Liu 2022), the multifunctional use of redox flow battery's electrolyte for data communication, in addition to powering, is novel, especially in the context of soft robotics. This work also describes a simple signal conditioning technique that addresses the signal decaying problem due to long-distance ionic data transmission. Finally, we describe the prototypical design of a decentralized data communication between two, or possibly more, systems of power and control. This work presents the concept of a decentralized control of a robot's hydraulic actuation system, that will allow for the system's functional robustness in the event of one of the two, or possibly more, modules lose power or become severed from the robot's body.

## II. Impedance characterization of tubing filled with electrolyte.

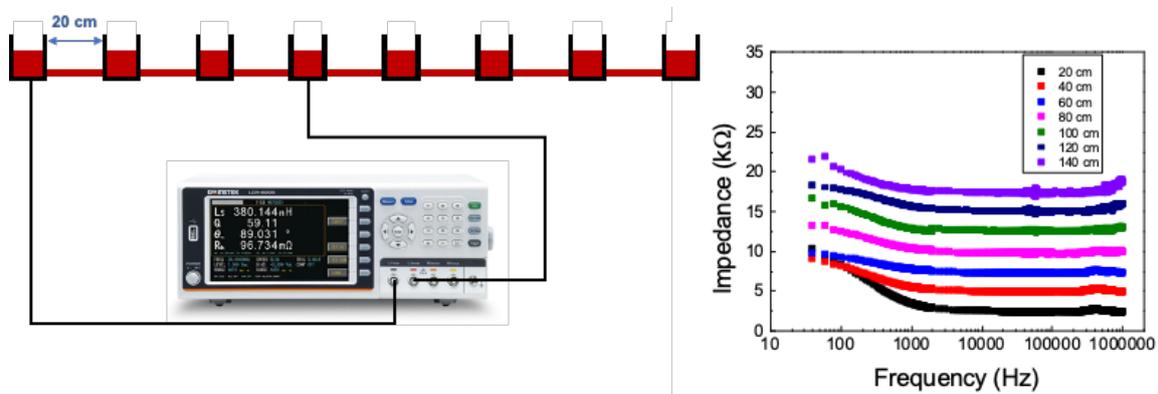

Fig. 1. Impedance of tube filled with electrolytes.

Based on the conductivity of the electrolyte, ions in the electrolyte vibrate upon application of a potential difference, making it suitable to be used as a conductive substrate to transfer data. When a signal is introduced by the microcontroller as an input, the ions in the electrolyte vibrate instantaneously due to the potential difference between the two ends of the electrolyte channel. Through vibration, the signal can be transmitted to the other end of the tube and read by the microcontroller there. The electrolyte has zinc iodide (3 M, solvent: deionized (DI) water with 10 wt% of ethanol) as ion components. For testing the data communication property via the electrolyte, 1.4 m-long (inner diameter: 5 mm) tube filled with the $ZnI_2$ electrolyte (concentration: 3 M) was prepared, as shown in Fig. 1. An impedance measurement was made every 20 Hz from 20 Hz to 2 kHz, every 500 Hz from 2 kHz to 100 kHz, and every 5 kHz from 100 kHz to 1 MHz. The impedance was saturated after 1 kHz and showed a constant value. With an increase in length from 20 cm to 140 cm, the impedance increased from 2 kΩ to 20 kΩ. As a result, an approximate impedance value of the electrolyte could be determined and used to match impedances in data communication systems.

## III. Signal conditioning for long-range data transmission via electrolytic channel

The transmitted data integrity between two microcontrollers over the two electrolyte channels can be improved by addressing the mismatched impedance at the receiving end of the transmission line, this is mainly due to the low input impedance of the digital input pin from the receiving microcontroller and the high impedance of the electrolyte that has been previously characterized, as shown in Fig. 2 A and B. As a solution, we connect the output of the electrolyte channel to the non-inverting input of a high-bandwidth op-amp (THS4304, Texas Instruments) that is set up to be a voltage buffer, as shown in Fig. 2C. The high input impedance of the voltage follower prevents the transmitted voltage from being distorted.

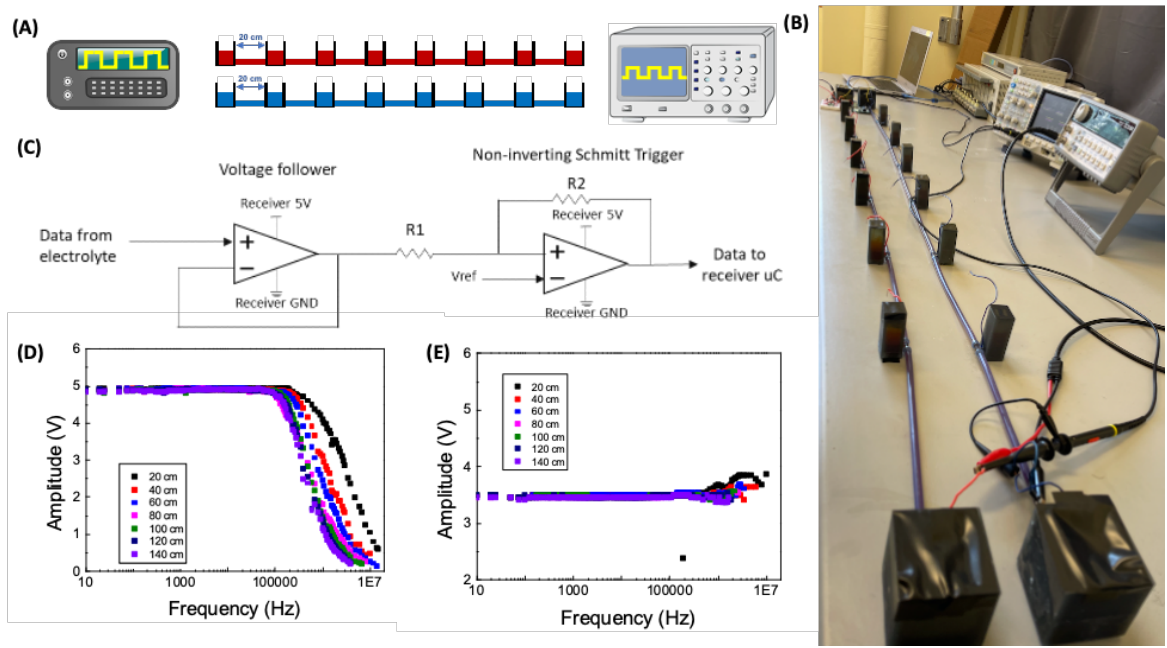

Fig. 2. One-way signal transmission measurement. (A) and (B) a schematic and photograph of the signal transmission test set-up. (C) Circuit diagram of the high-bandwidth op-amps with non-inverting Schmitt trigger arrangement. (D) Transferred signal without op-amps. (E) Transferred signal with op-amps.

For the second signal conditioning stage, we connect the output of the voltage buffer to a non-inverting Schmitt Trigger with the triggering input low and high thresholds to be 1.9 V and 3.2 V respectively. The output voltage rail-to-rail swings between 0.8 V low and 4.8 V high, that can be detected by the digital input pin of the receiving microcontroller. To achieve this Schmitt Trigger configuration, the resistors R1 and R2 values are 3.3 kΩ and 10 kΩ respectively, with the reference voltage of 2.5V.

As shown in Fig. 2D and E, when the signal transmission was measured by connecting the function generator and the oscilloscope using two electrolyte channels without an op-amp, the amplitude decreased at frequencies above 1 MHz. Furthermore, most of the amplitude was lost when it exceeded 10 MHz. However, by connecting the op-amp to the end of the channel and matching the impedance, the amplitude was maintained regardless of the increase in frequency.

To demonstrate the improvement of the communication over the electrolyte, we setup a one-way data transmission between two Arduino Unos using SoftwareSerial data communication with varied baud rates (bit rates) from 300 to 19,200 baud and varied electrolyte channel length from 20 to 140 cm with an increment of 20 cm. Due to the signal voltage of 5V from the transmitting microcontroller, there is hydrolysis reaction that degrade the electrolyte over time. This effect has not yet been studied in detail. We send a series of numbers from 0 to 7, with each number as the content of a data packet. Without the Schmitt Trigger and voltage buffer, the data packet is lost at 80 cm electrolyte length irrespective of the data baud rate. The data communication is significantly improved, without any packet lost, for any baud rate up to 19,200 baud and electrolyte channel length up to 140 cm when the voltage follower and Schmitt Trigger are implemented, as shown in Fig. 3.

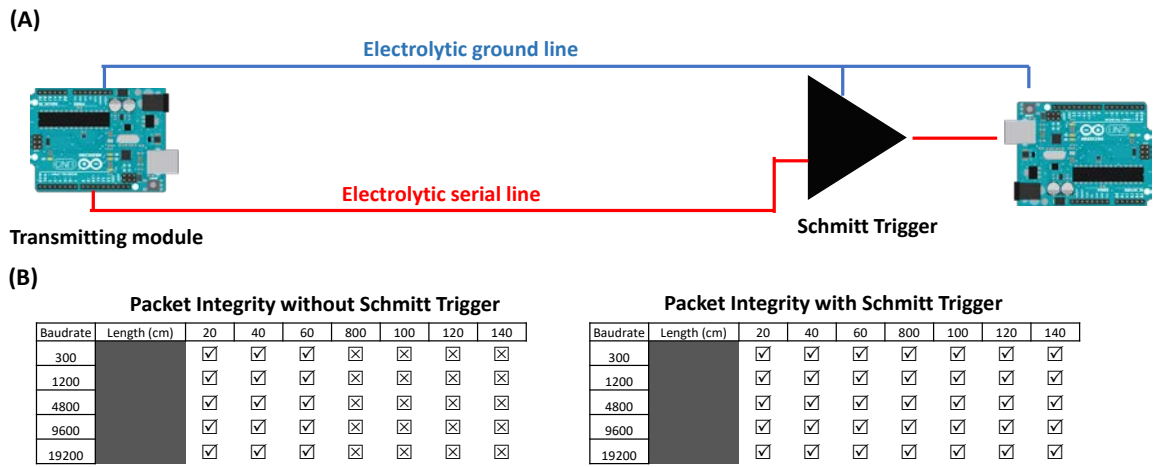

Fig. 3. One-way data transmission between two Arduino microcontrollers with different baud rates.

To evaluate how far the data transmission can reach with the aid of the Schmitt trigger, we demonstrate 10-meter-long one-way data transmission between a master Arduino Uno that is connected to a potentiometer and a slave Arduino Uno that is connected to a RGB ring LED (Neopixel 12 x 5050 RGB LEDs, Adafruit; Fig. 4). The potentiometer voltage divider value is mapped to the location (from 0 to 11) of the LEDs in the ring, thus allowing the user to control the LED location by turning the potentiometer knob. With the same electrolytic channel arrangement as shown in Fig. 3, the electrolytic ground line is connected to the GND pins of both Uno boards, and the electrolytic serial line is connected between a digital pin of the master board and IN+ pin of the Schmitt trigger. Fig. 4 shows that we can change the LED location by turning the potentiometer knob.

## IV. Two-way data communication via electrolytic channels for decentralized hydraulic control systems

This section demonstrates the two-way data communication via three electrolytic channels with potential applications for decentralized hydraulic control systems. Each control system consists of an electronic module and a set of two rechargeable Zinc-Iodide redox flow battery cells connected in series. The two identical flow battery cells, each having an open-circuit voltage of 1.2 V at 20% SOC, are connected in series to generate a total 2.4 V needed for the boost converter (u1v11a, Pololu). With 50 mA/cm$^2$ current density for each battery cell and the designed electrode area of 20 cm$^2$, each battery module delivers up to 100 mA. The battery module's charging and discharging profiles after 5 cycles can be seen in Fig. 5. Initially, both anolyte and catholyte colors are identical (light yellow), but after charging, the catholyte side becomes dark red. With approximately 124 mA current delivery (Fig. 5) and 2.2 V output, the 270 mW power delivery is sufficient for the electronic module's power consumption of 50 mW, even with 80% conversion efficiency from the booster. The electronic module consists of an Arduino Pro Mini 3.3 V, a boost converter to convert 2.2 V from the battery module to 3.3V that is required by the microcontroller, and an op-amp Schmitt trigger. The two modules establish common ground for data communication via the anolyte (yellow-colored) channel. These microcontrollers employ Software Serial data communication with one another at 9600 baud via two catholyte (dark red-colored) channels. The physical representation of the two communications modules can be seen in Fig. 6A and its corresponding schematic is shown in Fig. 6B.

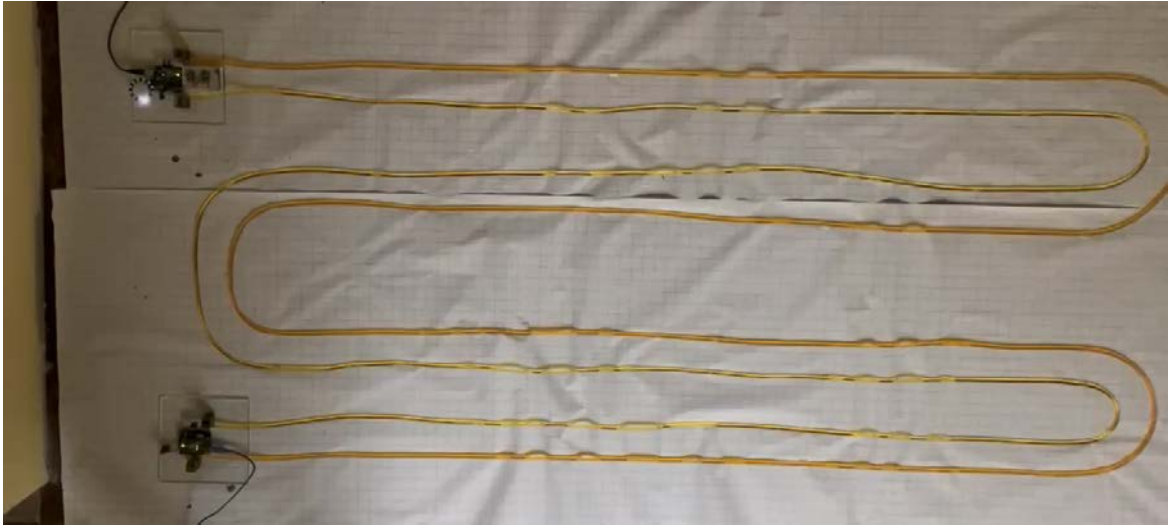

Fig. 4. Demonstration of data transmission over 10-meter-long electrolytic channels.

The two microcontrollers are programmed identically with two operating modes: one with established data communication and the other without it. After turning on two power switches of the two modules, the left microcontroller begins the LED blinking sequence, then sends a pulse, via TX line through the electrolyte channel, to the right microcontroller to notify the right microcontroller that the left side's LED sequence is complete. If the right microcontroller receives the pulse, it begins its own LED blinking sequence and sends back a pulse via the second electrolyte channel to the left microcontroller. This sequential data communication and LED blinking processes between the left and the right module continues until one of the two microcontrollers is disconnected from the power source. When this event happens, the remaining microcontroller switches to the other operating mode. In this mode, the microcontroller continues its LED blinking sequence continuously without waiting for any triggering event from external signals.

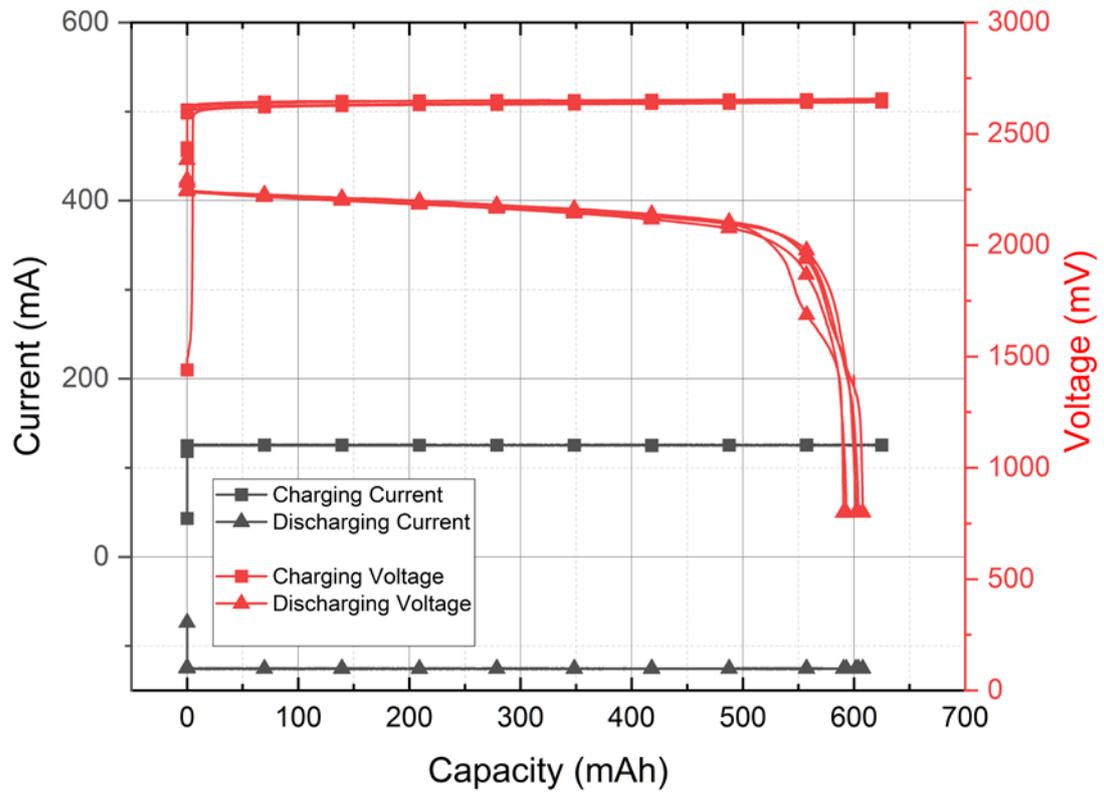

Fig. 5. Five cycles of charging and discharging of a Zinc – Iodide redox flow battery module that comprises of two battery cells in series.

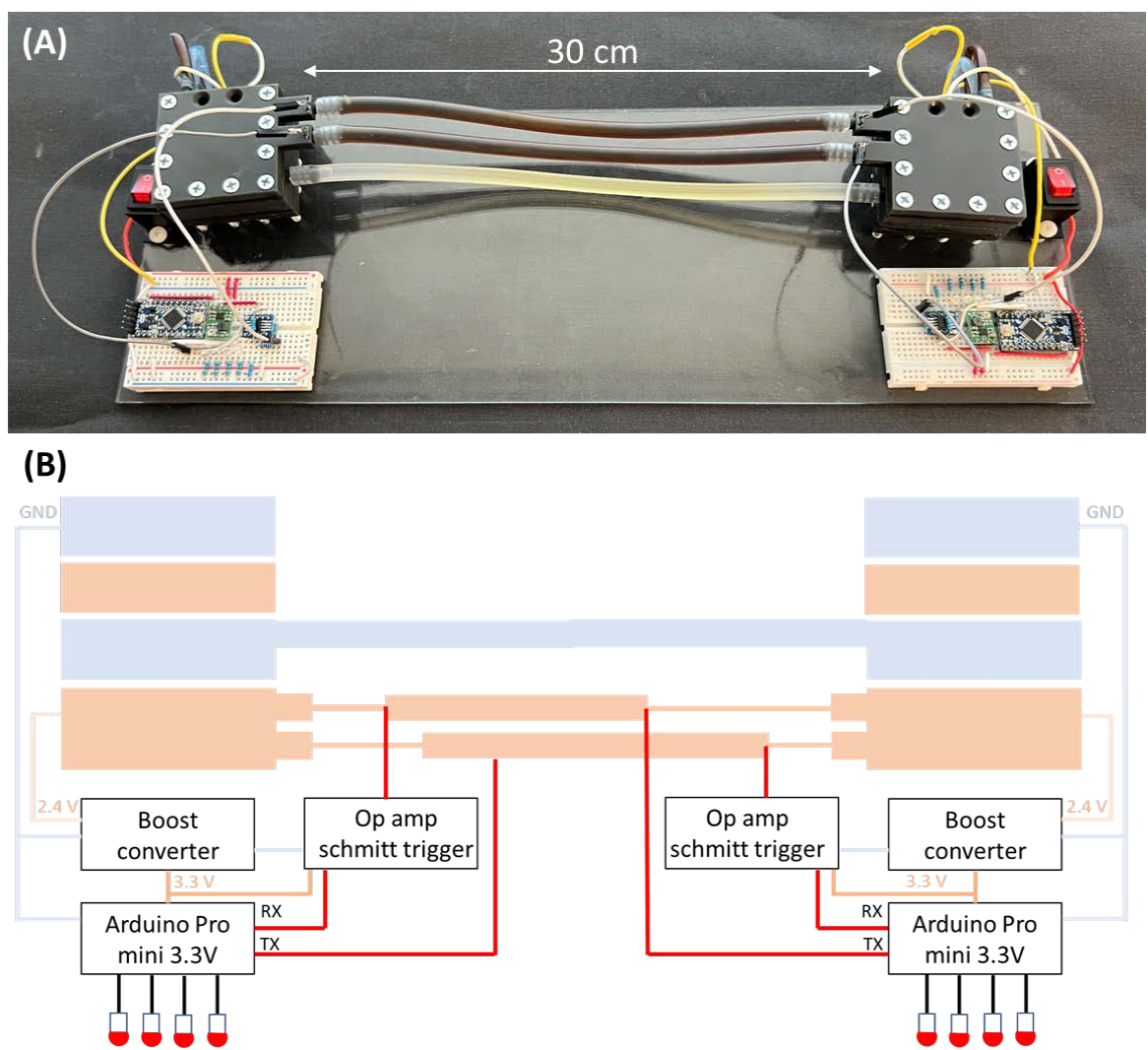

Fig. 6. Demonstration of two-way data communication between two Arduino Pro Mini 3.3V microcontrollers. The two systems are powered by two separated Zinc-Iodide redox flow battery modules. (A) Physical representations of the two communication modules that are connected to one another via three electrolytic channels. (B) The illustrated schematic of the systems.